\RequirePackage{fix-cm}
\documentclass[smallextended]{svjour3}       
\smartqed  
\usepackage{graphicx}
\usepackage{multirow}
\journalname{Advances in Data Analysis and Classification}
\usepackage{natbib,color}
\usepackage{amsfonts,amssymb,amsmath,amscd,latexsym,dsfont,url}

\newcommand{\bpi}{\boldsymbol{\pi}}
\newcommand{\by}{\mathbf{y}}
\newcommand{\bz}{\mathbf{z}}
\newcommand{\bV}{\mathbf{V}}
\newcommand{\clust}{\text{clust}}
\newcommand{\diff}{\text{diff}}
\newcommand{\indep}{\text{indep}}
\newcommand{\reg}{\text{reg}}
\newcommand{\clas}{\text{clas}}
\newcommand{\BIC}{\mbox{BIC}}
\newcommand{\crit}{\mbox{crit}}
\newcommand{\indi}{\mathds{1}}
\newcommand{\argmin}{\mathop{\mathrm{arg\,min}}}
\newcommand{\B}{{\mathcal B}}
\newcommand{\OO}{\mathcal O}

\begin{document}

\title{Variable selection in model-based clustering and discriminant analysis with a regularization approach}

\titlerunning{Variable selection in model-based clustering and classification}        

\author{Gilles Celeux \and Cathy Maugis-Rabusseau  \and Mohammed Sedki}


\institute{   Gilles Celeux \at
                 Inria and Universit\'e  Paris-Sud\\
  		Dept. de mathématiques\\
  		Bâtiment 425\\
  		91405 Orsay Cedex, France\\
                 \email{gilles.celeux@inria.fr}\\
           \and
                 Cathy Maugis-Rabusseau \at
                 INSA de Toulouse\\
  		135 avenue de Rangueil\\
 		 31077 Toulouse, Cedex 4, France\\
 		 \email{cathy.maugis@insa-toulouse.fr}\\
  \and
  		Mohammed Sedki \at
              	INSERM U$1181$ B2PHI, Institut Pasteur and UVSQ\\
  	     	B\^atiment. $15/16$ Inserm, H\^opital Paul Brousse\\
  	    	$16$ avenue Paul Vaillant Couturier\\
  	   	94807 Villejuif Cedex, France\\              
	   	\email{Mohammed.Sedki@u-psud.fr}          
}

\date{Received: date / Accepted: date}

\maketitle

\newpage
\begin{abstract}
Relevant methods of variable selection have been proposed in model-based clustering and classification. These methods are making use of backward or forward procedures to define the roles of the variables. Unfortunately, these stepwise procedures are terribly slow and make these variable selection algorithms inefficient to treat large data sets.  
In this paper, an alternative regularization approach of variable selection is proposed for
model-based clustering and classification.  In this approach, the variables are first ranked with a lasso-like procedure in order to avoid painfully slow stepwise algorithms. Thus, the variable selection methodology of   \citet{Maugis2009b} can be efficiently applied on high-dimensional data sets. 
\keywords{Variable Selection \and Lasso Procedure \and Gaussian Mixture \and Clustering \and
Classification}
\end{abstract}

\section{Introduction}
In data mining and statistical learning, available datasets are larger
and larger.  As a result, there is more and more interest in variable
selection procedures for clustering and classification tasks.  After a
series of papers on variable selection in model-based clustering
\citep{Law:04,Tadesse:05,Raftery:06,Maugis2009a}, \citet{Maugis2009b}
proposed a general model for selecting variables for clustering with
Gaussian mixtures.  This model, called SRUW, distinguishes between
relevant variables $(S)$ and irrelevant variables $(S^c)$ for
clustering.  In addition, the irrelevant variables are divided into
two categories.  A part of the irrelevant variables $(U)$ may be
dependent on a subset $R$ of the relevant variables and an other part
$(W)$ are independent of other variables.  In \citet{Maugis2009b}, a
procedure using embedded stepwise variable selection algorithms is
used to identify the SRUW sets. It leads to compare two models at each
step in order to determine which variable should be excluded or
included in the set $S$, $R$, $U$ or $W$. But these stepwise
procedures implemented in \textit{SelvarClustIndep}\footnote{\textit{SelvarClustIndep} is implemented in C++ and is available at \url{http://www.math.univ-toulouse.fr/~maugis/}} , are limited as soon as the number of variables is of the order of a few tens. The SRUW
model has been adapted to the Gaussian model-based classification
framework in \citet{Maugis2011}, see also \citet{Murphy:10}.  In this
supervised framework, the identification of the sets $S$, $R$, $U$ and
$W$ is simpler since the stepwise procedures are not performed inside
an EM algorithm but the stepwise algorithms still encounters
combinatorial explosion phenomenons.

In parallel \citet{Pan2007} were inspired by the success of the lasso
regression to develop a method of variable selection in model-based
clustering using $\ell_1$ regularization of the likelihood. This
approach was successively extended in \citet{Xie2008,Wang2008} and
finally \citet{Zhou2009} proposed a regularized Gaussian mixture model
with unconstrained covariance matrices.  

In the present paper, the variables are first ranked using a $\ell_1$ penalty  placed on the Gaussian mixture component mean vectors and precision matrices.  This is made
feasible by exploiting the abundant literature on lasso penalization
in Gaussian graphical models \citep[see][]{Friedman2007, Meinshausen2006}.  
Using the resulting ranking of the variables
avoids combinatorial problems of the stepwise variable selection algorithms. And, it is hoped that using this lasso-like
ranking of the variables instead of stepwise algorithms would not
deteriorate the identification of the sets $S$, $R$, $U$ and $W$.

This article is organized as follows. In Section~\ref{sect-SRUW}, the SRUW model is
reviewed in the Gaussian model-based clustering context and its simple
extension to the Gaussian model-based classification context is
sketched.  The variable selection procedure using lasso-like
penalization is presented in Section~\ref{sect-lassochinois} focusing on model-based clustering. Section~\ref{Sect:numericalexp} is devoted to the comparison of the procedures \textit{SelvarClustIndep} and \textit{SelvarMix}\footnote{\textit{SelvarMix} \textsf{R} package is available on cran.} on several simulated datasets. 
A short discussion section ends the article.

\section{The SRUW model}\label{sect-SRUW}
\subsection{Model-based clustering}\label{subsect-modelbasedclutering}
Let $n$ observations $\by=(\by_1,\ldots,\by_n)'$ be described by $p$
continuous variables ($\by_{i}\in\mathbb{R}^p$). In the model-based
clustering framework, the multivariate continuous data $\by$ are
assumed to come from several subpopulations (clusters) modeled with a
multivariate Gaussian density. The observations are assumed to arise
from a finite Gaussian mixture with $K$ components and a mixture form
$m$, namely
$$
  f\big(\by_i \mid K,m,\alpha\big)=\underset{k=1}{\stackrel{K}{\sum}} \pi_k
  \phi\big(\by_i\mid \mu_k,\Sigma_{k(m)}\big),
$$
where $\bpi=(\pi_1,\ldots,\pi_K)$ is the mixing proportion vector
($\pi_k\in(0,1)$ for all $k=1,\ldots,K$ and $\sum_{k=1}^K \pi_k=1$),
$\phi\big(\cdot \mid\mu_k,\Sigma_k\big)$ is the $p$-dimensional
Gaussian density function with mean $\mu_k$ and covariance matrix
$\Sigma_k$, and
$\alpha=\big(\bpi,\mu_1,\ldots,\mu_K,\Sigma_1,\ldots,\Sigma_K\big)$ is
the parameter vector. Several Gaussian mixture forms $m$ are
available, each corresponding to different assumptions on the forms of
the covariance matrices, arising from a modified spectral
decomposition.  These include whether the volume, shape and
orientation of each mixture component vary between components or are
constant across clusters
\citep{BanfieldRaftery1993,CeleuxGovaert1995}.

Typically, the mixture parameters are estimated via maximum likelihood
using the EM algorithm \citep{Dempster:77}, and both the number of
components $K$ and the mixture form $m$ are chosen using the Bayesian
Information Criterion (BIC) \citep{Schwarz:78} or other penalized
likelihood criteria as the Integrated Completed Likelihood (ICL)
criterion \citep{Biernacki:00} in the clustering context.  Among the
R packages which implement this methodology, we could
mention the \textit{mclust}~\citep{Scrucca2016} software,
and the \textit{Rmixmod}~\citep{Lebret2015} software. 

\subsection{Variable selection in model-based clustering}
The SRUW model, as described in \citet{Maugis2009b}, involves three
possible roles for the variables: the relevant clustering variables
($S$), the redundant variables ($U$) and the independent variables
($W$).  Moreover, the redundant variables $U$ are explained by a
subset $R$ of the relevant variables $S$, while the variables $W$ are
assumed to be independent of the relevant variables.  Thus the data
density is assumed to be decomposed into three parts as follows:
\begin{eqnarray*}
& & f(\by_i \mid K,m,r,\ell,\bV,\theta) = \\
& & \underset{k=1}{\stackrel{K}{\sum}}
  \pi_k \phi\left(\by_i^S \mid \mu_k,\Sigma_{k(m)}\right) 
  \times \phi\left(\by_i^U \mid a+\by_i^R b,\Omega_{(r)}\right) 
  \times\phi\left(\by_i^W \mid \gamma,\tau_{(\ell)}\right)
\end{eqnarray*}
where $\theta=(\alpha,a,b,\Omega,\gamma,\tau)$ is the full parameter
vector and $\bV = (S,R,U,W)$.  The form of the regression covariance
matrix $\Omega$ is denoted by $r$; it can be spherical, diagonal or
general.  The form of the covariance matrix $\tau$ of the independent
variables $W$ is denoted by $\ell$ and can be spherical or diagonal.

The SRUW model recasts the variable selection problem for model-based
clustering as a model selection problem, where the model collection is
indexed by $(K,m,r,\ell,S,R,U,W)$. This model selection problem is
solved maximizing the following BIC-type criterion:
\begin{eqnarray}
& & \crit \big(K,m,r,\ell,\bV\big) = \nonumber\\
& &\BIC_{\clust}\left(\by^S \mid K, m\right)
  + \BIC_{\reg}\left(\by^U \mid r, \by^R\right)
  + \BIC_{\indep}\left(\by^W \mid \ell\right),\label{eq:Crit}
\end{eqnarray}
where $\BIC_{\clust}$ represents the BIC criterion of the Gaussian
mixture model with the variables $S$, $\BIC_{\reg}$ represents the BIC
criterion of the regression model of the variables $U$ on the
variables $R$ and $\BIC_{\indep}$ represents the BIC criterion of the
Gaussian model with the variables $W$.

Since the SRUW model collection is large, two embedded backward or
forward stepwise algorithms for variable selection, one for the
clustering and one for the linear regression, are considered to solve
this model selection problem.  A backward algorithm allows one to
start with all variables in order to take variable dependencies into
account. A forward procedure, starting with an empty clustering
variable set or a small variable subset, could be preferred for
numerical reasons when the number of variables is large.  The method is
implemented in the \textit{SelvarClustIndep} software and a simplified version is implemented in the \textit{clustvarsel}\footnote{\textit{clustvarsel} \textsf{R} package is available on cran.} \textsf{R} package.
But in a high-dimensional setting, even the variable selection method with the
two forward stepwise algorithms becomes painfully slow and alternative
methods are desirable.

\subsection{Variable selection in Classification}
In the supervised classification framework, the labels of the training
dataset are known and the variable selection problem is analogous but
simpler than in the clustering framework.  The training data set is
$$
(\by,\bz)=\{(\by_1,z_1),\ldots,(\by_n,z_n);\ \by_i\in\mathbb{R}^p, z_i\in\{1,\ldots,K\}\},
$$
where $\by_i$ is the $p$-dimensional i.i.d predictors and $z_i$ are  the 
corresponding class labels. The number of classes $K$ is
known. The subset $S$ is now the discriminant variable subset.  Under
a model $(m,r,\ell,\bV)$ with $\bV=(S,R,U,W)$, the distribution of the
training sample is modeled by
$$
\left\{\begin{array}{r c l}
    f(\by_i|z_i=k,m,r,\ell,\bV)& = & \phi(\by_i^S|\mu_k,\Sigma_{k(m)})\, \phi(\by_i^U|a+\by^R\beta,\Omega_{(r)})\, \phi(\by_i^W|\gamma,\tau_{(\ell)})\\
    \\
    (\indi_{z_i=1},\ldots,\indi_{z_i=K}) & \sim &
    \mbox{Multinomial}(1;\pi_1,\ldots,\pi_K)
  \end{array}.\right.
$$
According to the assumed form of the covariance matrices involving
their eigenvalue decomposition, a collection of more or less
parsimonious mixture forms $(m)$ is available as in the clustering context.

Considering the same model SRUW, the variable selection problem is
solved by using the following model selection criterion
$$
  \crit(m,r,\ell,\bV) = \BIC_{\clas}(\by^S,\bz|m)
  + \BIC_{\reg}(\by^U|r,\by^R)+ \BIC_{\indep}(\by^W|\ell),
$$
where $\BIC_{\clas}$  denotes the BIC criterion for the Gaussian
classification on the discriminant variable subset $S$.  Maximizing this criterion is an
easier task in this supervised context since there is no need to use
the EM algorithm to derive the parameter estimates of the Gaussian
classification model contrary to the model-based clustering
situation. See~\citet{Maugis2011} for details. But, the variable selection procedure with two embedded
stepwise procedures remain expensive and alternative procedures are
desirable.

\section{Variable selection through regularization}\label{sect-lassochinois}
In order to avoid the highly CPU-time consuming of stepwise algorithms, 
we propose an alternative variable selection procedure in two steps: First, the variables are ranked through a
lasso-like procedure, and second, the variable roles are determined
using criterion~\eqref{eq:Crit} on these ranked variables. The
procedure is detailed in the clustering framework, the simplifications
for classification are presented in
Section~\ref{subsection-algo-DA}. This variable selection procedure is
implemented in the R package \textit{SelvarMix}.

\subsection{Variable ranking by regularization}~\label{Subsect:variableranking}

In the first step, the variables are ranked through the lasso-like procedure of
\citet{Zhou2009}.  For any $K\in\mathbb{N}^\star$, the criterion to be maximized is
\begin{equation}\label{crit-chinois}
 \sum_{i=1}^n \ln \Big[ \sum_{k=1}^K \pi_k \phi\big(\bar{\by}_i \mid \mu_k,
  \Sigma_k\big)\Big] - \lambda \sum^K_{k = 1} \left\|\mu_k\right\|_1
  - \rho \sum_{k = 1}^K \left\|\Sigma_k^{-1}\right\|_1,
\end{equation}
where
$$\left\|\mu_k\right\|_1 = \underset{j=1}{\stackrel{p}{\sum}} \big|\mu_{kj}\big|,
\left\|\Sigma_k^{-1}\right\|_1 = \underset{j^\prime \neq j}{\sum_{j^\prime =1}^p\sum_{j = 1}^p} \left|(\Sigma_k^{-1})_{jj^\prime}\right|,$$
$\bar{\by}_i=(y_{ij}- \bar{\by}_j)_{1\leq j\leq p}$ with $\bar{\by}_j =
\frac 1 n \sum_{i=1}^n y_{ij}$ and where $\lambda$ and $\rho$ are two
non negative regularization parameters defined on two grids of values
$\mathcal G_\lambda$ and $\mathcal G_\rho$ respectively. The estimated
mixture parameters for fixed tuning parameters $\lambda$ and $\rho$,
$$\widehat{\alpha}(\lambda, \rho) = \big(\widehat{\bpi}(\lambda, \rho),
\widehat{\mu}_1(\lambda, \rho), \ldots, \widehat{\mu}_K(\lambda,
\rho), \widehat{\Sigma}_1(\lambda, \rho), \ldots,
\widehat{\Sigma}_K(\lambda, \rho)\big)$$ are computed with the EM
algorithm of \citet{Zhou2009}. In particular, the glasso
algorithm \citep{Friedman2007} using a coordinate descent procedure
for the lasso is used to estimate the sparse precision matrices
$\Sigma_k^{-1}, k = 1, \ldots, K$. This procedure is reminded in Appendix
\ref{Appendix:lassoclustering}.

It is worth noting that this lasso-like criterion does not take into
account the typology of the variables induced by the SRUW
model. Strictly speaking, it only distinguishes two possible roles for
the variables: a variable is declared related or independent of the
clustering.  A variable is declared independent for the clustering if for all
$j=1,\ldots,p$, and $k=1,\ldots, K$, $\widehat{\mu}_k(\lambda,
\rho)=0$. The variance matrices are not considered in this definition.
Actually, their role is secondary in clustering and taking them into account would imply
serious numerical difficulties.

Varying the regularization parameters $(\lambda, \rho)$ in $\mathcal
G_\lambda\times \mathcal G_\rho$, a "clustering" score is defined for each variable
$j\in\{1,\ldots,p\}$ and for fixed $K$ by
$$
  \OO_K(j)=\sum_{(\lambda, \rho) \in \mathcal G_\lambda \times \mathcal G_\rho}  \B_{(K, \rho, \lambda)}(j)
$$
where
$$
  \B_{(K, \rho, \lambda)} (j) =
  \begin{cases}
    0 & \mbox{if } \widehat{\mu}_{1j}(\lambda,\rho) = \ldots = \widehat{\mu}_{Kj}(\lambda,\rho) = 0 \\
    1 & \mbox{else}.
  \end{cases}
$$
The larger $\OO_K(j)$ is the more related to the clustering the variable $j$ is expected to be.
The variables are thus ranked by their decreasing values on $\OO_K(j)$. This variable ranking is denoted
$\mathcal I_K =(j_1,\ldots,j_p)$ with $\OO_K(j_1)> \OO_K(j_2)>\ldots>\OO_K(j_p)$.

\subsection{Determination of the variable roles}\label{Subsect:cluster-role}
\begin{figure}[h]
  \centerline{\includegraphics[width=8cm]{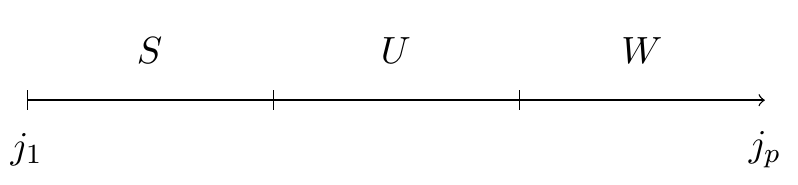}}
  \caption{The ideal position of the sets $S$, $U$ and $W$ in the ranking of the variables}
  \label{fig:Order}
\end{figure}

The relevant clustering variable set $S_{(K,m)}$ is determined
first. The variable set is scanned according to the $\mathcal I_K$
order.  One variable is added to $S_{(K,m)}$ if
\begin{eqnarray} 
    \BIC_\diff(j_v) &= &\BIC_\clust\Big(\by^{S_{(K,m)}},\by^{j_v} \mid K,m\Big)\nonumber\\
                    &  & -  \BIC_\clust\Big(\by^{S_{(K,m)}} \mid K,m\Big) - \BIC_{\reg}\Big(\by^{j_v} \mid \by^{R[j_v]}\Big)\label{eq:BICdiff}
\end{eqnarray}
is positive, $R[j_v]$ being the variables of $S_{(K,m)}$ required to
linearly explain $\by^{j_v}$. This subset $R[j_v]$ is determined with
standard backward stepwise algorithm for variable selection in linear regression. The scanning of
$\mathcal I_K$ is stopped as soon as $c$ successive variables have a
non positive $\BIC_\diff$ value, $c$ being a fixed positive integer.
Once the relevant variable set $S_{(K,m)}$ is determined, the
independent variable set $W$ is determined as follows. Scanning the
variable set according to the reverse order of $I_K$, a variable $j_v$
is added to $W_{(K,m)}$ if the subset $R[j_v]$ of $S_{(K,m)}$ (derived
from the backward stepwise algorithm) is empty. The algorithm stops as
soon as $c$ successive variables are not declared independent. The
redundant variables are thus declared to be $U_{(K,m)}=\{1,\ldots,p\}
\setminus \{S_{(K,m)}\cup W_{(K,m)}\}$ and the subset $R_{(K,m,r)}$ of
$S_{(K,m)}$ required to linearly explain $\by^{U_{(K,m)}}$ is derived
from the backward stepwise algorithm, for each covariance shape $r$.
The ideal position of the variable sets $S$, $U$ and $W$ in the
variable ranking is schematised in Figure~\ref{fig:Order}.  Finally,
the model $(\hat K,\hat m,\hat r, \hat \ell)$ maximizing the criterion
$\crit \big(K,m,r,\ell,\bV_{(K,m,r)}\big)$ defined in
Equation~\eqref{eq:Crit} with $ \bV_{(K,m,r)} =
(S_{(K,m)},R_{(K,m,r)},U_{(K,m)},W_{(K,m)})$ is selected.

Some comments are in order:
\begin{itemize}
\item It is possible to use a lasso procedure instead of the stepwise
  variable selection algorithm in the linear regression step.
  However, this replacement is not expected to be highly beneficial
  since stepwise variable selection in linear regression are not too
  much expensive and, moreover, the number of variables in the set $S$
  is not expected to be high.
  \item There is no guarantee that the variable order designed in
    \ref{Subsect:variableranking} would be in accordance with the
    ideal ranking of the variables displayed in Figure
    \ref{fig:Order}. In particular when the variables are highly
    correlated, lasso-like procedures could be expected to produce
    confusion between the sets $S$ and $U$. This is the reason why we
    wait $c\ (>1)$ few steps before deciding the variable roles: we
    give a chance to the procedure to catch more variables in $S$ and
    in $W$.
\end{itemize}

\subsection{Variable selection through regularization for classification}\label{subsection-algo-DA}
In the classification context, $K$ is fixed and the regularization criterion to be maximized is
\begin{equation}\label{crit-chinois-DA}
  \sum_{i = 1}^n\sum_{k=1}^K \indi_{\{z_i=k\}} \ln \Big[\pi_k \phi\big(\bar{\by}_i \mid \mu_k,
  \Sigma_k\big)\Big]- \lambda \sum^K_{k = 1} \left\|\mu_k\right\|_1
  - \rho \sum_{k = 1}^K \left\|\Sigma_k^{-1}\right\|_1,
\end{equation}
with the same notation as in Section~\ref{Subsect:variableranking}.
Assuming that the training data set has been obtained according to the
mixture sampling scheme, the proportions $\pi_1, \ldots, \pi_K$ are
estimated by
\begin{equation*}
  \widehat{\pi}_k = \frac 1 n \sum_{i = 1}^n \indi_{\{z_i = k\}}, \ k=1,\ldots,K.
\end{equation*}
The maximization of criterion~\eqref{crit-chinois-DA} is done
according to the procedure described in Appendix~\ref{Appendix:lassoclassif}.
The only difference with the
clustering context is that the labels $z_i$ are known and no EM
algorithm is required.  Thus, a ranking $\mathcal I_K$ of the
variables is get and the procedure described in
Section~\ref{Subsect:cluster-role} for model-based clustering is
adapted straightforwardly to the supervised classification context,
where \eqref{eq:BICdiff} is replaced by

\begin{eqnarray*}
& &  \BIC_\diff(j_v) = \\
& & \BIC_{\clas}\left(\by^{S_{(m)}},\by^{j_v},\bz \mid m\right)
  -  \BIC_{\clas}\left(\by^{S_{(m)}},\bz \mid m\right)
  - \BIC_{\reg}\left(\by^{j_v} \mid \by^{R[j_v]}\right).
\end{eqnarray*}

\section{Numerical experiments}\label{Sect:numericalexp}

This section is devoted to comparing our procedure implemented in the R package \textit{SelvarMix}
with the forward/backward stepwise procedures of \citet{Maugis2009b, Maugis2011} in both model-based clustering and classification settings.

\subsection{Model-based clustering}\label{Subsect-NumExp-clustering}
\subsubsection{Comparison on simuated data}
We consider one of the seven simulated data sets studied in
\citet[Section~6.1]{Maugis2009b}. The data consist of $n=2000$
observations on $p=14$ variables.  On the first two variables
($S=\{1,2\}$), data are distributed from an equiprobable mixture of
four Gaussian distributions $\mathcal N (\mu_k, I_2)$ with $\mu_1 =
(0,0), \mu_2=(4,0), \mu_3 = (0,2)$ and $\mu_4 = (4,2)$.  On the nine
redundant variables ($U=\{3, \ldots, 11\}$), data are simulated as
follows: for $i=1,\ldots,n$,
\begin{equation*}
    \by_i^{\{3-11\}}  = (0, 0, 0.4, 0.8, \ldots,2) + \by_i^{\{1,2\}} b + \varepsilon_i
 \end{equation*}
where the regression coefficients are
\begin{equation*}
  b = \big((0.5,1)^\prime,(2,0)^\prime, (0,3)^\prime,(-1, 2)^\prime,
  (2,-4)^\prime, (0.5, 0)^\prime, (4, 0.5)^\prime, (3,0)^\prime, (2,1)^\prime\big)
\end{equation*}
and $\varepsilon_i$ are i.i.i $\mathcal N(0_9, \Omega)$. The
regression covariance matrix $\Omega$ is block diagonal
$$\Omega= \text{diag}\big(I_3, 0.5I_2,\Omega_1, \Omega_2\big)$$ 
with $\Omega_1 =
\text{Rot}\big(\frac \pi 3\big)^\prime \text{diag}(1, 3)
\text{Rot}\big(\frac \pi 3\big)$ and $\Omega_2 = \text{Rot}\big(\frac
\pi 6\big)^\prime \text{diag}(2, 6) \text{Rot}\big(\frac \pi 6\big)$,
where $\text{Rot}(\theta)$ is a plane rotation matrix with angle
$\theta$. The last three independent variables are standard Gaussian
random variables $\by_i^{\{12-14\}} \sim \mathcal N \left(0_3,
  I_3\right)$.

\begin{figure}[htbp]
\centerline{
  \begin{tabular}{c}
  \includegraphics[width=8cm]{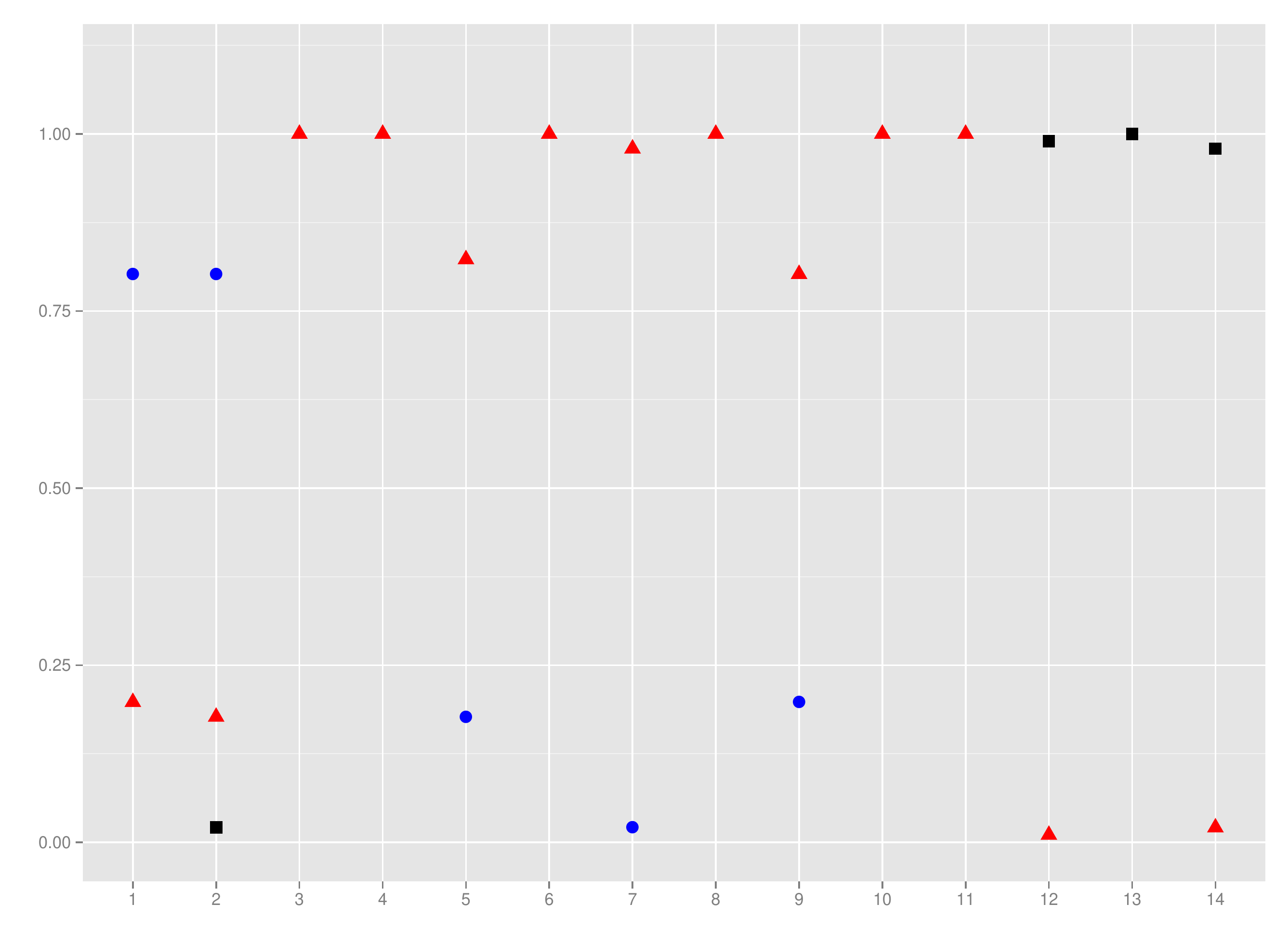}\\
  \includegraphics[width=8cm]{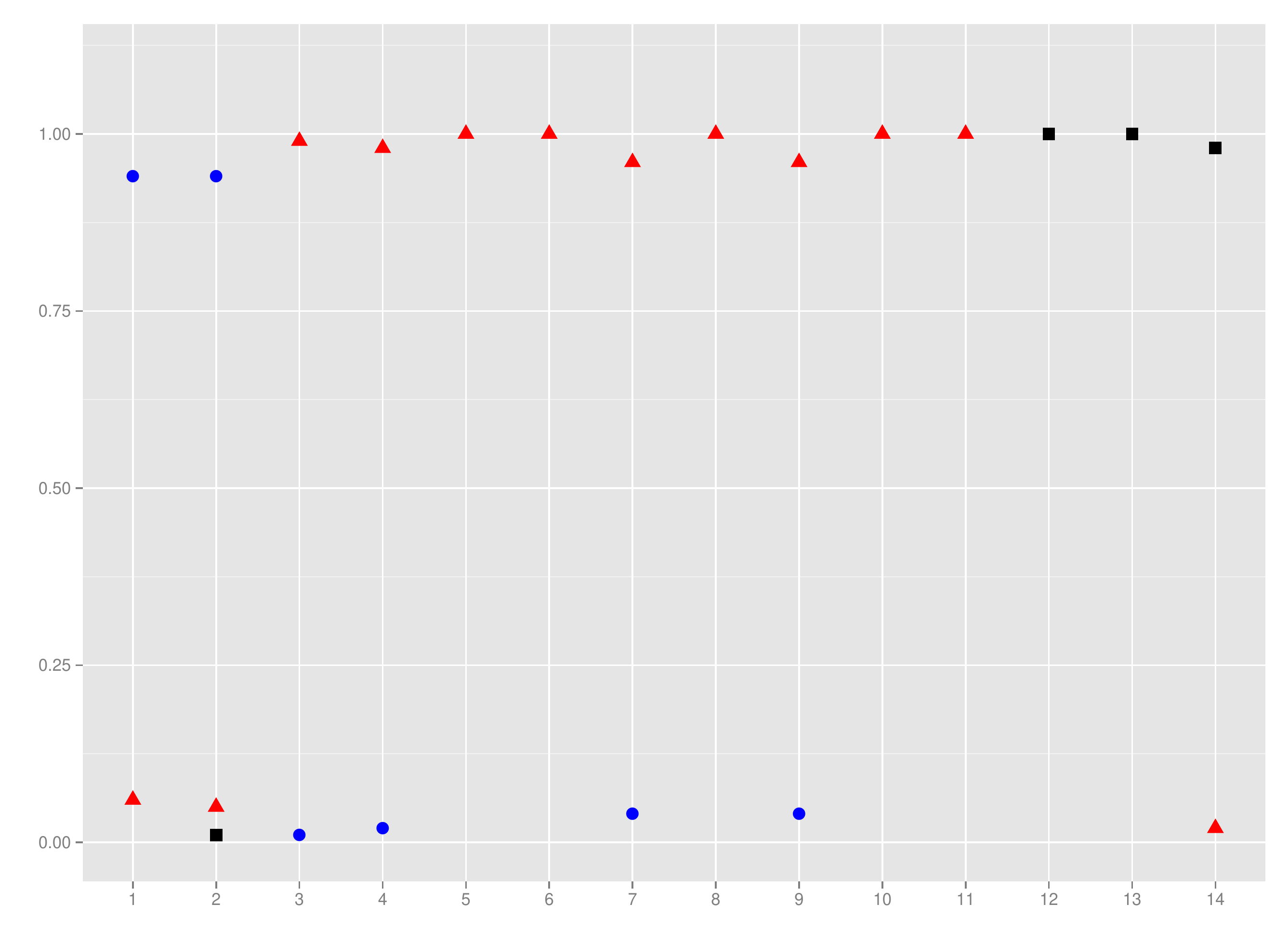}
\end{tabular}}
  \caption{Proportion of times each variable was declared
           relevant (square), redundant (triangle) or independent
           (circle) by \textit{SelvarClustIndep} (top) and \textit{SelvarMix}
           (bottom) in the first scenario. Zero values are not shown.
           \label{fig:FreqNonSup}}
\end{figure}

In the first scenario, the function \textit{SelvarClustLasso} of
\textit{SelvarMix} and the forward \citet{Maugis2009b}'s procedure
(\textit{SelvarClustIndep}) are compared on $100$ replications of the
simulated dataset with $K=4$ and spherical mixture
components. We compare the CPU times of both
  procedures. The calculations are carried out on an 80 Intel Xeon
  2.4 GHz processors machine and the variable ranking procedure (see
  Section~\ref{Subsect:cluster-role}) of \textit{SelvarMix} is
  parallelized.  In comparaison to \textit{SelvarMix}, the combinatorial
  nature of \textit{SelvarClustIndep} makes it difficult to be parallelized.
  As a result, a significant improvement of the runtime
  with \textit{SelvarMix} is obtained: \textit{SelvarMix} takes $47.0 (\pm 3.2)$    seconds CPU time whereas \textit{SelvarClustIndep} needs $450.64(\pm 104.0)$. Figure~\ref{fig:FreqNonSup} displays the distribution of the variable
  roles with \textit{SelvarClustIndep} in left and \textit{SelvarMix} in
  right. Globally, the true variable roles are well recovered.
  Surprisingly \textit{SelvarMix} detects the relevant variables better
  than  \textit{SelvarClustIndep} which sometimes selects variables $5$
  and $9$ instead of the first two variables.

\begin{figure}[htbp]
\centerline{
  \begin{tabular}{c}
  \includegraphics[width=8cm]{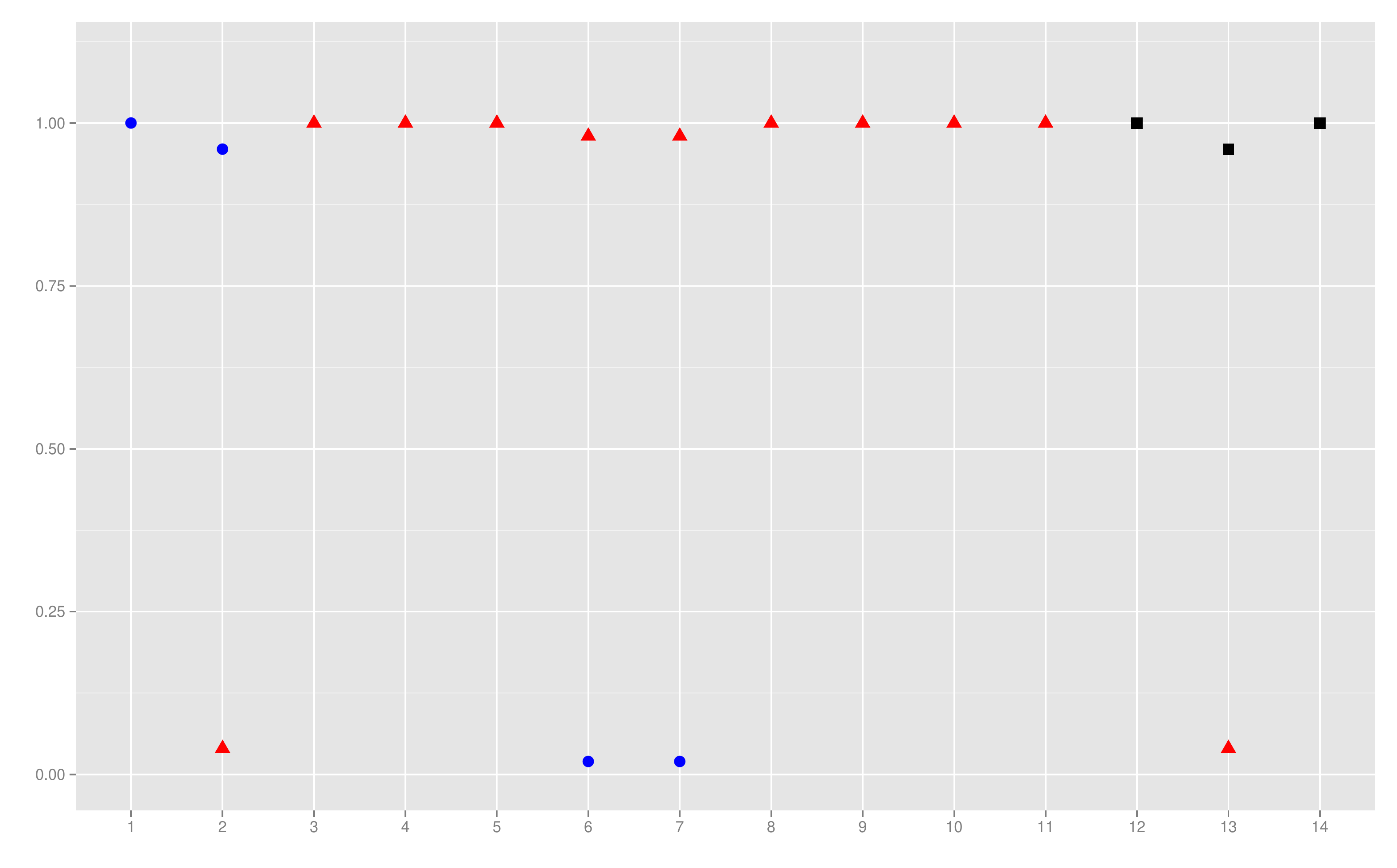}
  \\
  \includegraphics[width=8cm]{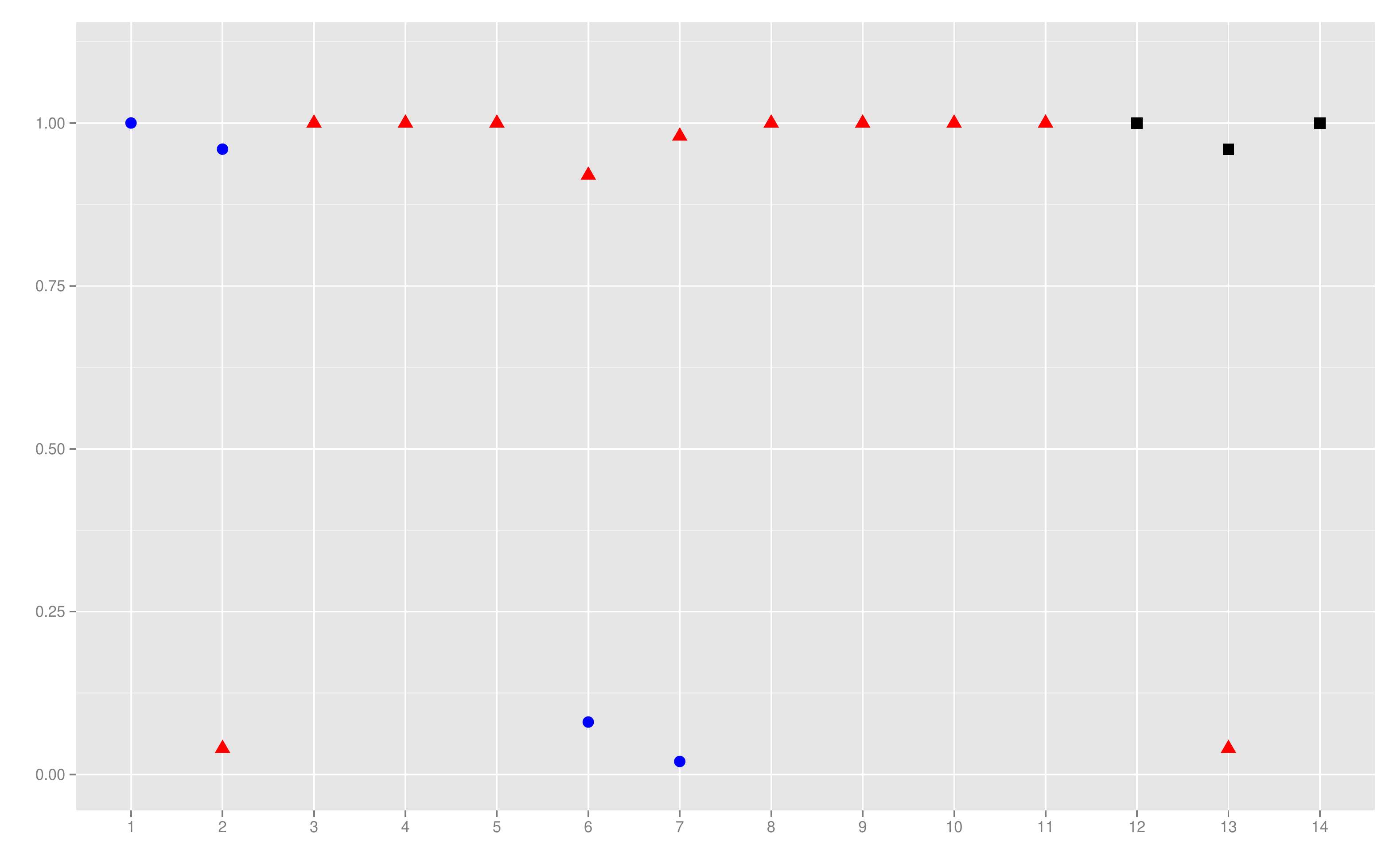}
\end{tabular}}
  \caption{Proportion of times each variable was declared relevant (square),
           redundant (triangle) or independent (circle) by \textit{SelvarClustIndep}
           (top) and \textit{SelvarMix} (bottom) in the second scenario.
           Zero values are not shown.\label{fig:FreqNonSup2}}
\end{figure}

In the second scenario, the 50 previous simulated datasets are
considered but now the number of clusters $K$ varies between 2 to 6
and the 28 Gaussian mixture forms $m$ are considered.  The true
cluster number is always correctly selected by both
procedures. The variable selection is similar with
both procedures (see Figure~\ref{fig:FreqNonSup2}): the
true variable partition is selected 46 (resp. 48) times by
\textit{SelvarClustIndep} (resp. \textit{SelvarMix}). The clustering
performance is preserved with \textit{SelvarMix} since the average of
adjusted rand index (ARI) is $0.6 (\pm 0.017)$ with
\textit{SelvarMix} and $0.6 (\pm 0.015)$ with \textit{SelvarClustIndep}.

We also applied the \cite{Zhou2009}'s lasso-like procedure on the
$50$ simulated datasets. As previously, the number of clusters varies
from $2$ to $6$. The penalization parameters, including the means 
$\widehat{\mu}_1, \ldots, \widehat{\mu}_K$, are estimated according to 
the penalized likelihood criterion~\eqref{crit-chinois} whereas the number of 
clusters $K$ is selected using the BIC criterion. One variable $j$ is declared
relevant if there is at least one cluster $k$ where
$|\widehat{\mu}_{kj}| > 10^{-1}$.  For all $50$ simulated datasets, the \cite{Zhou2009}'s
procedure fails to select the true number of clusters ($K = 4$) and
the true set of relevant variables ($S = \{1,2\}$). It
always selects $K = 6$, both relevant and redundant
variables are declared relevant and the ARI is lower ($0.45 (\pm 0.025)$).

\begin{figure}[htbp]
\centerline{\includegraphics[width=\textwidth]{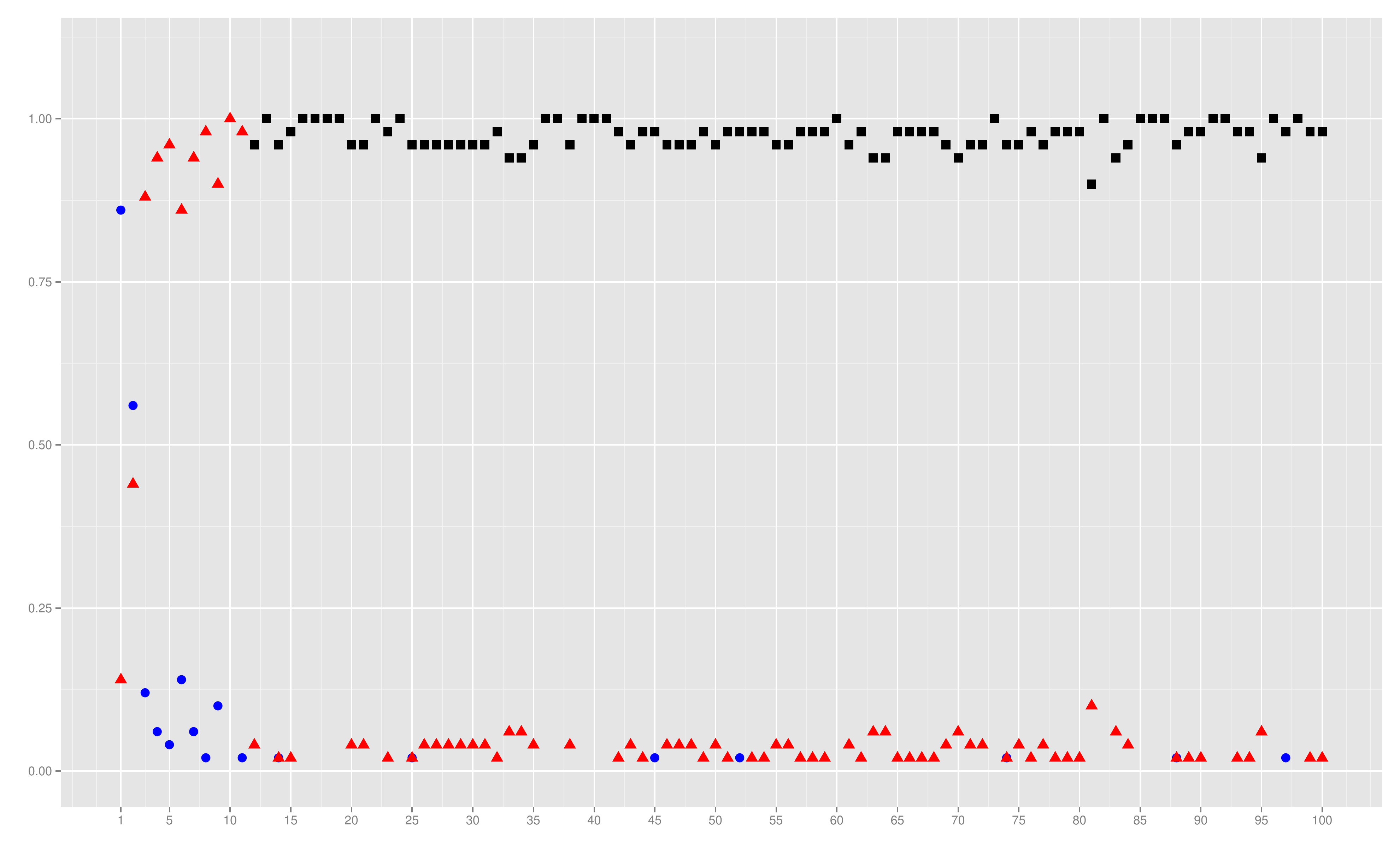}}
\caption{Proportion of times each variable was declared discriminant
        (square), redundant (triangle) or independent (circle) by
        \textit{SelvarMix} in the clustering setting, for the datasets
        with $p=100$ variables. Zero values are not shown.
        \label{fig:FreqCSDA100}}
\end{figure}

In the third scenario, we consider $50$ datasets consisting of $n=400$
observations described by $100$ variables. On the first eleven
variables, data are distributed as previously. Next, 89 standard
Gaussian variables are appended. As previously, the number of clusters
varies from $2$ to $6$ and the $28$ Gaussian mixtures forms $m$ are
considered. The selection of the number of clusters is less efficient: the
true cluster number $K = 4$ is selected $23$ times, the model with $K
= 3$ is selected $10$ times whereas the models with $K = 2, 5$ and $6$
are respectively selected $6, 7$ and $4$ times.  
Compared to the previous scenario in low dimension ($p = 14$),
the variable selection using \textit{SelvarMix} is slightly deteriorated
(see Figure~\ref{fig:FreqCSDA100}). The true relevant variable set $S = \{1,2\}$ is selected $23$ times.  
Sometimes, \textit{SelvarMix} declares one of the redundant variables as relevant. 
One of the independent variables is declared as relevant
seven times. Moreover, the clustering performance is slightly deteriorated: the ARI is $0.49 (\pm 0.1)$. 

\subsubsection{Comparison on real datasets}
In this section, we compare our proposal with the forward/backward stepwise procedure 
of \citet{Maugis2009b, Maugis2011} on the
moderate high-dimensional  data sets  waveform and transcriptome summarized in 
Table~\ref{datasets}.  Results and CPU times are shown in Table~\ref{wavetrans}.
\begin{table}[th!]
\begin{center}
\begin{tabular}{|cccccc|}
\hline 
 data               & $n$    & $d$   & covariance model & number of components & reference\\ 
  \hline         
waveform        & $5000$ & $41$  &   20 models      & from $3$ to $6$ & \cite{Maugis2009b}\\
transcriptome  & $4616$ & $33$  &  2 models        & from $10$ to $30$ &   \cite{Gagnot2008}\\
\hline
\end{tabular} 
\end{center}
\caption{Characteristics, references and experiment conditions for both data sets.\label{datasets}} 
\end{table}

\begin{table}[ht!]
\begin{center}
\color{black}
\begin{tabular}{|cc|c c c c c c c |c|}
\hline
data set & Software & BIC  & $\widehat{K}$ & $\widehat{m}$  & $\text{Card}(S)$
& $\text{Card}(R)$ & $\text{Card}(U)$ & $\text{Card}(W)$ & time\\
\hline
\multirow{2}{*}{waveform} & SelvarClustIndep  & $-591542$  & $6$  & pkLC  & $16$  & $8$  & $3$  & $21$  &  $>$ 24hrs\\
                          & SelvarMix  & $-602965$  & $6$  &  pLkC & $15$  & $11$  & $8$ & $18$  & $1.29$ mns\\
                          \hline
\multirow{2}{*}{transcriptome} & SelvarClustIndep  & $-694455$  & $24$ & pkLkC  & $30$  & $16$   & $3$  & $0$  & $>$ 24 hrs   \\
                          & SelvarMix  & $-695534$  & $27$  & pkLkC  & $29$  & $15$  & $4$  & $0$  & 1.49 hrs \\
 \hline
\end{tabular} 
\end{center}

\caption{Results and cpu-time for both methods on both data sets.\label{wavetrans}}
\end{table}
Table~\ref{wavetrans} shows that for the transcriptome data set the selected partitions have not the same number of clusters. However, their ARI ($0.60$) is high for a such large number of clusters. Moreover the sets $S,R$ and $U$ are almost identical. Since the CPU time of SelvarMix is dramatically smaller than SelvarClustIndep, this procedure can be preferred.

\begin{table}[ht]
\begin{center}
\begin{tabular}{ccccccc}
 & \multicolumn{6}{c}{SelvarClustIndep}\\
  \multirow{6}{*}{SelvarMix} & 606 &   0  &  0  &  0  &  0 &   1\\
   &  0 & 986  & 55  &  0  &  0 &  24\\
   &  1 &   0  &576  &  1  &  0 &   0\\
   & 22 &   0  & 19  &956  &  0 &   0\\
   & 61 &   0  &  0  &  0  &1063&   36\\
   &  1 &   0  &  1  &  0  &  0 & 591\\
\end{tabular}
\end{center}
\caption{Contingency table of SelvarClustIndep partition vs.  SelvarMix partition on the waveform data set. As shown by the table, both partitions are quite similar. The ARI is $0.90$.}
\end{table}

\subsection{Supervised classification}
We consider the same simulated example of \citet{Maugis2011}. The
samples are described by $p=16$ variables. The prior probabilities of
the four classes are $\bpi=(0.15, 0.3, 0.2,0.35)$. On the three
discriminant variables, data are distributed from 
$$\by_i^{\{1-3\}} \mid z_i = k \sim \mathcal N\left(\mu_k, \Sigma_k\right)$$ 
with means vectors 
{$\mu_1 = (1.5, -1.5, 1.5)$}, 
$\mu_2 = (-1.5, 1.5, 1.5)$, 
$\mu_3 =(1.5, -1.5, -1.5)$ and 
$\mu_4 = (-1.5, 1.5, -1.5)$. 
The covariance matrices are $\Sigma_k = \left(\rho_k^{|i-j|}\right)$ with
$\rho_1=0.85, \rho_2 = 0.1, \rho_3 = 0.65$ and $\rho_4 = 0.5$. There
are four redundant variables simulated from
$
  \by_i^{\{4-7\}} \sim \mathcal N \left( \by_i^{\{1,3\}}\beta , I_4\right).
$
with 
$$ \beta = \left(
   \begin{array}{cccc}
    1 &  0 & -1 & 2\\
    0 & -2 &  2 & 1
   \end{array}
  \right).
$$
 Nine independent variables are appended, sampled from
$\by^{\{8-16\}}_i \sim \mathcal N(\gamma, \tau)$ with
$$\gamma = (-2, -1.5, -1, -0.5, 0, 0.5, 1, 1.5, 2)$$
and the diagonal matrix
$$
\tau = \text{diag}(0.5, 0.75, 1, 1.25, 1.5, 1.25, 1, 0.75, 0.5).
$$

\begin{figure}[htbp]
  \centerline{
  \begin{tabular}{c}
  \includegraphics[width=8cm]{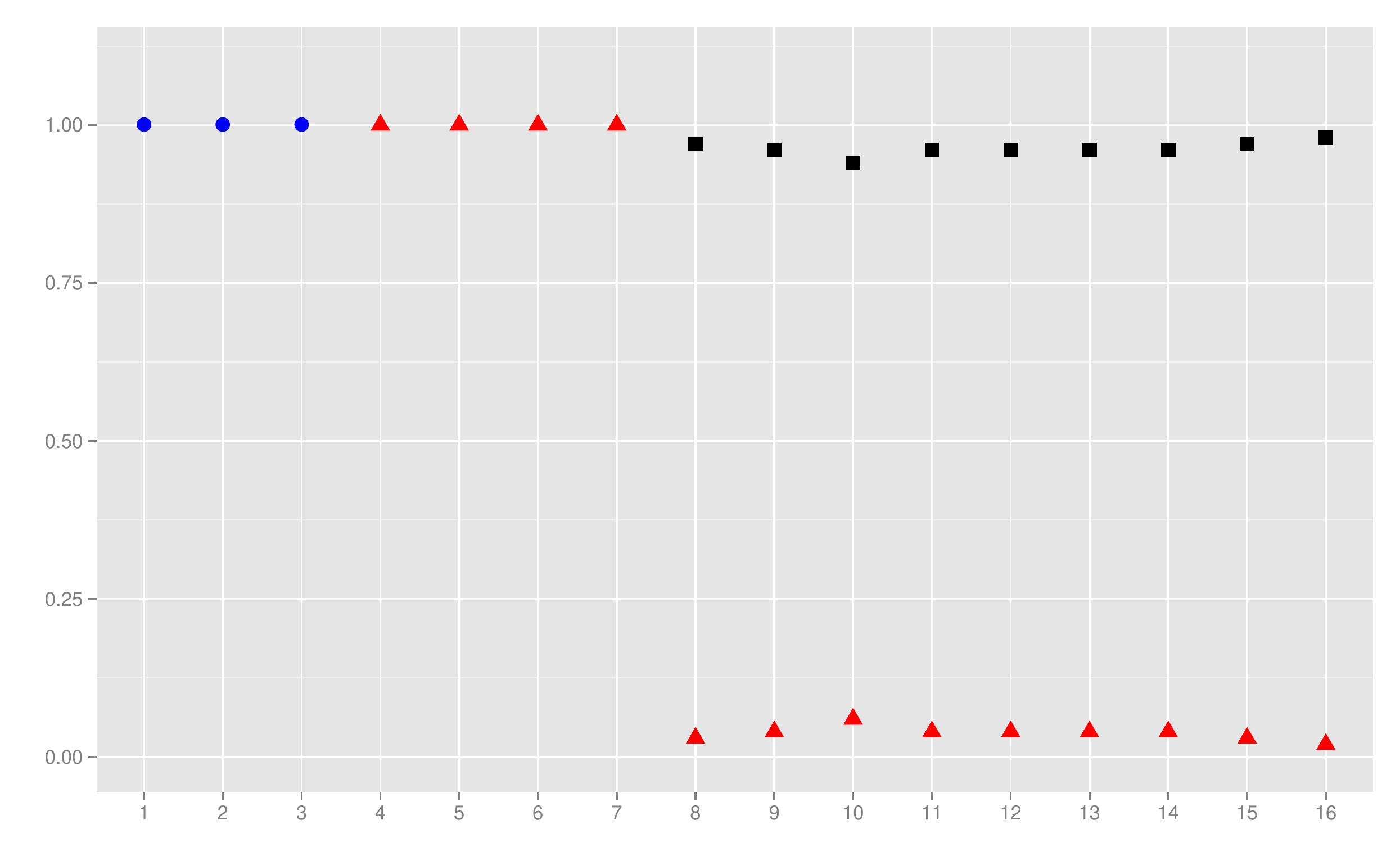}
  \\
  \includegraphics[width=8cm]{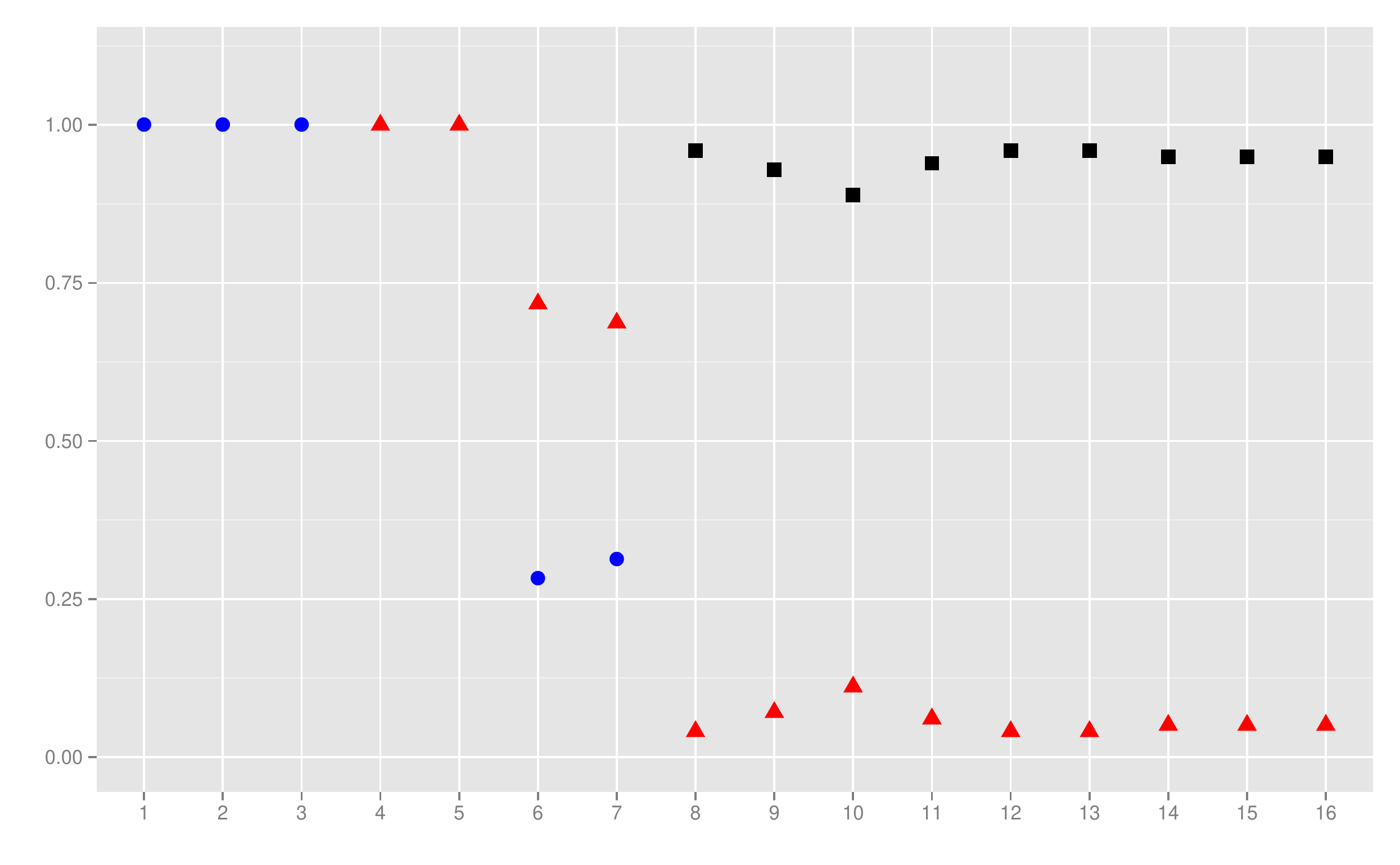}
\end{tabular}}
\caption{Proportion of times each variable was declared discriminant
        (square), redundant (triangle) or independent (circle) by
        \textit{SelvarClustIndep} (top) and \textit{SelvarMix} (bottom)
        in the classification setting, for the data sets with $p=16$ variables.
        Zero values are not shown. \label{fig:FreqSup}}
\end{figure}

First, $100$ simulation replications are considered where the training
sample is composed of $n=500$ observations and the same test sample
with $50\,000$ points is used.  The fourteen forms $m$ are
considered. The two procedures \textit{SelvarMix} and the version of
\textit{SelvarClustIndep} devoted to the variable selection in supervised
classification (which is again called SelvarClustIndep in the sequel)
are compared. Figure~\ref{fig:FreqSup} shows the variable selection
obtained with the procedure of \citet{Maugis2011} on the left and the
\textit{SelvarLearnLasso} function of \textit{SelvarMix} on the
right. \textit{SelvarMix} sometimes declares Variables 6 and 7 as
relevant with the first three relevant variables and has some tendency
to declare redundant some independent variables, more often than
\textit{SelvarClustIndep}. For the prediction, both procedures have a
similar behaviour: the misclassification error rate is $4.5 \% (\pm
0.19)$ for \textit{SelvarMix} and $4.18 \% (\pm 0.06)$ for
\textit{SelvarClustIndep}.

\begin{figure}[htbp]
\centerline{\includegraphics[width=\textwidth]{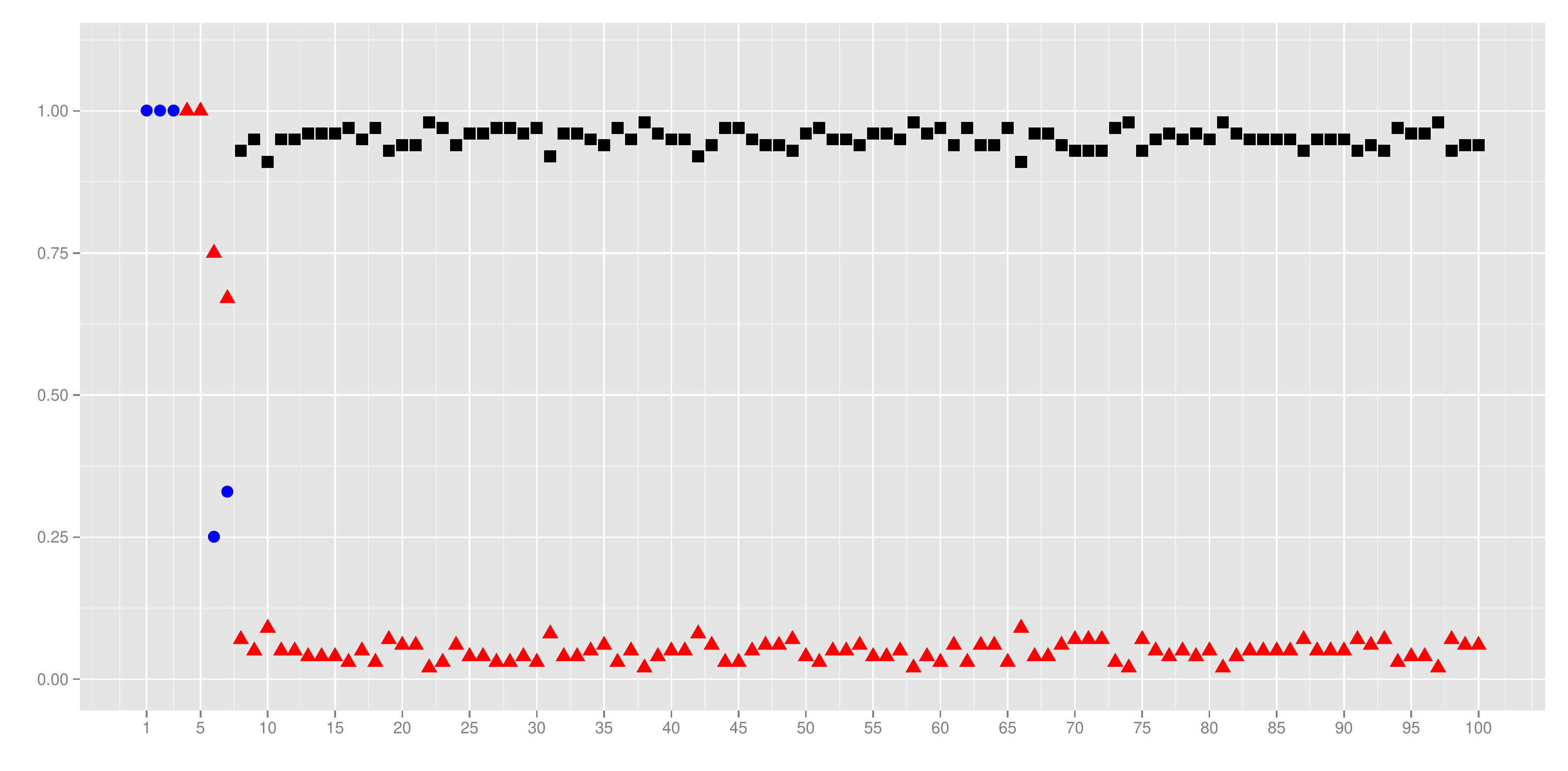}}
\caption{Proportion of times each variable was declared discriminant
         (square), redundant (triangle) or independent (circle) by
         \textit{SelvarMix} in the classification setting, for the data sets
         with $p=100$ variables. Zero values are not shown.
         \label{fig:FreqSupp100}}
\end{figure}

Second, we consider $100$ training samples composed of $n=500$
observations described by $p=100$ variables, $84$ standard Gaussian
variables are appended to the previous training samples.  The test
sample is similarly modified. As expected, \textit{SelvarMix} allows us
to quickly study such data sets where the number of variables is
large. The level of prediction is preserved since the
misclassification error rate is $4.34 \% (\pm 0.18)$ and the variable
selection remains similar as shown in
Figure~\ref{fig:FreqSupp100}.

\section{Discussion}
The model SRUW is a powerful model to define the roles of variables in
the Gaussian model-based clustering and classification
contexts. However the diffusion of this model is slowed down by the
stepwise selection algorithms used in its previous implementations.
Despite they are sub-optimal these algorithms are highly CPU time
demanding.

The regularization approach we propose allows us to avoid these
stepwise procedures by designing an irreversible order on which the
variables in $S$, $U$ and $W$ are chosen. The numerical experiments
performed with the resulting R package \textit{SelvarMix} show quite
encouraging performances. \textit{SelvarMix} is highly faster than
\textit{SelvarClustIndep} while providing analogous (and sometimes
better) performances than the reference \textit{SelvarClustIndep}
program.

In practice the influence of the number $c$ for which the
regularization algorithm provides the same role for $c$ successive
variables could be regarded as an important tuning parameter. In our
numerical experiments, this parameter appears to be not too much
sensitive and a default value $c = 3$ seems reasonable. However the
influence of $c$ needs to be further investigated in order to propose
heuristic rules to choose it as a simple function of the total number
of variables in a proper way to get stable selections.

\appendix
\section{Procedures to maximize penalized empirical contrasts}
\subsection{The model-based clustering case}\label{Appendix:lassoclustering}

The EM algorithm for maximizing Criterion~\eqref{crit-chinois} is as
follows \citep{Zhou2009}.  The penalized complete loglikelihood of the
centered data set $\bar{\by} = \big(\bar{\by}_1, \ldots, \bar{\by}_n
\big)'$ is given by
\begin{equation}
  \label{eq:CompletedLikelihood}
  L_{\text{c},(\lambda, \rho)} (\bar{\by}, \bz, \alpha) =
  \sum_{i =1}^n \sum_{k = 1}^K z_{ik} \big[ \ln(\pi_k) + \ln\phi(\bar{\by}_i \mid \mu_k, \Sigma_k)\big] - \lambda \sum^K_{k = 1} \left\|\mu_k\right\|_1 - \rho \sum_{k = 1}^K
  \left\|\Theta_k\right\|_1
\end{equation}
where $\Theta_k=\Sigma_k^{-1}$ denotes the precision matrix of the
$k$-th mixture component.  The EM algorithm of \citet{Zhou2009}
maximizes at each iteration the conditional expectation of
\eqref{eq:CompletedLikelihood} given $\bar{\by}$ and a current
parameter vector $\alpha^{(s)}$: $\mathbb E\Big[L_{\text{c},(\lambda,
  \rho)}\big(\bar{\by}, \bz, \alpha \big) \mid \bar{\by},
\alpha^{(s)}\Big].$ The following two steps are repeated from an
initial $\alpha^{(0)}$ until convergence. At the $s$-th iteration of
the EM algorithm:
\begin{itemize}
\item[$\bullet$] \textbf{E-step:}
  The conditional probabilities $t^{(s)}_{ik}$ that the $i$-th
  observation belongs to the $k$-th cluster are computed for
  $i=1,\ldots,n$ and $k=1,\ldots,K$,
  $$
  t^{(s)}_{ik} = \mathbb{P}\big(z_{ik} = 1 \mid \bar{\by}, \alpha^{(s)}\big)
  = \frac{\pi_k^{(s)} \phi\Big( \bar{\by}_i \mid \mu^{(s)}_k,\Sigma^{(s)}_k \Big)}{ \sum_{k' = 1}^K \pi_{k'}^{(s)} \phi\Big(\bar{\by}_i \mid \mu^{(s)}_{k'}, \Sigma^{(s)}_{k'} \Big)}.
  $$

\item[\bf $\bullet$] \textbf{M-step :} This step consists of
  maximizing the expected completed loglikelihood derived from the
  E-step. It leads to the following mixture parameter updates:
  \begin{itemize}
  \item The updated proportions are $\pi^{(s+1)}_k = \frac 1 n \sum_{i = 1}^n t^{(s)}_{ik}$ for $k=1,\ldots,K$.

  \item Compute the updated means $\mu^{(s+1)}_1, \ldots,
    \mu^{(s+1)}_K$ using the formulas $(14)$ et $(15)$ of
    \citet{Zhou2009}: the $j$-th coordinate of $\mu^{(s+1)}_k$ is the
    solution of the following equations
    $$
    \text{if}  \quad \left| \sum_{i = 1}^n t^{(s)}_{ik} \left[\underset{v \neq j}{\sum_{v = 1}^p}
        \left(\bar{\by}_{ij} - \mu^{(s)}_{k v}\right) \Theta^{(s)}_{k,v j} + \bar{\by}_{ij}
        \Theta^{(s)}_{k,jj}\right]\right| \le \lambda
    \quad
    \text{then}
    \quad
    \mu^{(s+1)}_{kj} = 0,
    $$
    otherwise
    \begin{align*}
      \left[\sum_{i = 1}^n t^{(s)}_{i k}\right] \mu^{(s+1)}_{kj}
      \Theta^{(s)}_{k,jj} + \ \lambda \ \text{sign}\left(
        \mu^{(s+1)}_{kj}\right)
      & = \sum_{i = 1}^n t_{ik}^{(s)} \sum_{v = 1}^p \bar{y}_{iv} \Theta^{(s)}_{k,v j} \\
      & - \left[\sum_{i = 1}^n t^{(s)}_{i k}\right] \left[\left(\sum_{v =
            1}^p \mu^{(s)}_{kv} \Theta^{(s)}_{k,vj}\right) -
        \mu^{(s)}_{kj} \Theta^{(s)}_{k, jj}\right].
    \end{align*}

  \item For all $k=1,\ldots,K$, the covariance matrix
    $\Sigma_k^{(s+1)}$ is obtained via the precision matrix
    $\Theta_k^{(s+1)}$. The \textit{glasso} algorithm \citep[available in
    the R package \textit{glasso} of][]{glasso} is used to solve the
    following minimization problem on the set of symmetric positive
    definite matrices (denoted $\Theta \succ 0$):
        $$
          \underset{\Theta \succ 0}{\argmin}\left\{ -\ln \text{det}\left(\Theta \right) + \text{trace}\left(S_k^{(s+1)} \Theta\right) + \rho_k^{(s+1)} \|\Theta\|_1\right\},
        $$
where $ \rho_k^{(s+1)} = 2 \rho \left(\sum_{i = 1}^n t^{(s)}_{ik}\right)^{-1}$ and
    $$
    S^{(s+1)}_k = \frac{\sum_{i = 1}^n t^{(s)}_{ik}(\bar{\by}_i - \mu^{(s+1)}_k) (\bar{\by}_i - \mu^{(s+1)}_k)^\top}{\sum_{i = 1}^n t^{(s)}_{ik}}.
    $$
  \end{itemize}
\end{itemize}

\subsection{The classification case}\label{Appendix:lassoclassif}
The maximization of the regularized criterion \eqref{crit-chinois-DA}
at $\mu_1, \ldots, \mu_K$ and $\Theta_1, \ldots, \Theta_K$ is achieved
using an algorithm similar to the one presented in
Section~\ref{Appendix:lassoclustering} when the labels $z_i$ are
known.

The $j$-th coordinate of the mean vector $\mu_k$ is the solution of
the following equations
\begin{equation*}
  \text{if}  \quad \left| \sum_{i = 1}^n \indi_{\{z_i = k\}} \left[
      \underset{v \neq j}{\sum_{v = 1}^p} \left(\bar{\by}_{ij} -
        \mu_{k v}\right) \Theta_{k,v j} + \bar{\by}_{ij}
      \Theta_{k,jj}\right]\right| \le \lambda
  \quad
  \text{then}
  \quad
  \mu_{kj} = 0,
\end{equation*}
otherwise
\begin{align*}
  \left[\sum_{i = 1}^n \indi_{\{z_i = k\}} \right] \mu_{kj} \Theta_{k,jj} + \lambda\ \text{sign}\left( \mu_{kj}\right)
  & = \sum_{i = 1}^n \indi_{\{z_i = k\}} \sum_{v = 1}^p \bar{y}_{iv} \Theta_{k,v j} \\
  & - \left[\sum_{i = 1}^n \indi_{\{z_i = k\}} \right] \left[\left(\sum_{v = 1}^p \mu_{kv} \Theta_{k,v j}\right) - \mu_{kj} \Theta_{k, jj}\right].
\end{align*}

To estimate the sparse precision matrices $\Theta_1, \ldots, \Theta_K$
from the data set $\by$ and the labels $\bz$, we use the glasso
algorithm to solve the following minimization problem on the set of
symmetric positive definite matrices
\begin{equation}
  \label{eq:PenalizedLogLikDA}
  \widehat{\Theta}_k = \underset{\Theta \succ 0}{\argmin}\Big\{ -\ln
  \text{det}\big(\Theta \big) +  \text{trace}\big(S_k \Theta\big) + \rho_k \|\Theta\|_1\Big\},
\end{equation}
for each $k = 1, \ldots, K$. The $\ell_1$ regularization parameter in
\eqref{eq:PenalizedLogLikDA} is given by
$\rho_k = 2 \rho \left(\sum_{i = 1}^n \indi_{\{z_{i} = k\}}\right)^{-1}$
and the empirical covariance matrix $S_k$ is given by
\begin{equation*}
  S_k = \frac{\sum_{i = 1}^n \indi_{\{z_i=k\}}  (\bar{\by}_i -\mu_k) (\bar{\by}_i - \mu_k)^\top }{ \sum_{i =1}^n\indi_{\{z_i=k\}} }.
\end{equation*}

And, a coordinate descent maximization in $(\mu_1, \ldots, \mu_K)$ and
$(\Theta_1, \ldots, \Theta_K)$ is achieved until
convergence.

\section*{Acknowledgments}
One of the authors was supported by Fondation de Coop\'eration
Scientifique Campus Paris-Saclay-DIGITEO during his one year post-doc
at INRIA Saclay \^Ile-de-France.  This work was partially
supported by the French Agence Nationale de la Recherche (ANR), under
grant MixStatSeq (ANR-13-JS01-0001-01).

\bibliographystyle{spbasic}      
\bibliography{biblio}   

\end{document}